# The Nature of the Hadronic String

*UKQCD Collaboration*

**C. Michael**

DAMTP, University of Liverpool, Liverpool, L69 3BX, U.K.

**and**

**P. W. Stephenson**

Department of Physics, University College of Swansea, Swansea, SA2 8PP, U.K.

**Abstract**

We study a closed hadronic string using pure glue lattice simulation. We measure the energy of the flux state encircling the periodic boundary condition (the torelon). From these data we deduce the string fluctuation component and compare with models for the hadronic string.

## 1  Introduction

The quantum theory of a string was motivated by the observation that hadronic systems had string-like features. Now that we can evaluate hadronic properties from first principles using lattice gauge theory simulation, it is possible to explore the status of this string model. The string is intended to model the gluon flux between colour sources. There is copious evidence that this is a good approximation when the colour sources are sufficiently far apart. Since the string model appears to be reasonable, it is worthwhile to quantify the nature of this string. This can be explored by looking at the zero point fluctuations which result in energy shifts of the string energy spectrum. This is the topic which we address in this paper.

The fluctuating string is taken as the description of the colour flux between coloured sources (i.e. quarks). In lattice calculations, static colour sources can be used so an appropriate method to explore the string model is from a study of the long range static potential $V(R)$. As well as the ground state potential $V(R)$, one may study excited states of the gluon field between static sources. These have been determined by lattice simulation and compared to the predictions of a string model. The agreement is good at large $R$ [1] where the string model is expected to apply.

However, the quantum nature of the string is only revealed by study of the string fluctuation term. As is well known [2], the string fluctuation component in $V(R)$ for a bosonic string behaves as $\pi/(12R)$ at large $R$. Suggestions have been made that other self-energy expressions are appropriate for different string models. A Dirac string [3] leads to a self-energy term four times smaller. Thus it would be useful to determine the self-energy contribution accurately from lattice calculations. Unfortunately, the $1/R$ behaviour makes it difficult to separate this unambiguously from the Coulombic component which also has a $1/R$ behaviour. Precisely this observation lead to the proposal [4] to explore 3 dimensional gauge theories where the Coulomb and string fluctuation effects are clearly different ($\log R$ and $1/R$ respectively).

One should try to avoid these Coulombic contributions. One way to separate effects due to gluon exchange (Coulombic) and string fluctuations (confining force) is from an analysis of the spin-orbit potentials. This gives further information since the two possible spin-orbit potentials are related [5] to the central potential by $V = V_2 - V_1$. Furthermore, the Coulombic component contributes to $V_2$ dominantly so that a measurement of $V_1$ allows the long-range confining force to be studied with Coulombic contributions suppressed. Lattice determinations can separate these two components [6] and they indicate that, at the moderate $R$ values ($R < 0.3$ GeV$^{-1}$) studied, there is no significant sign of a string fluctuation piece in $V_1$. More accurate results which extend to larger $R$ are needed to reinforce any conclusion from this approach.

Another way forward is possible from a study of closed strings. These are systems with colour flux encircling the periodic spatial boundary of size $L$. The energy $E(L)$ of such a torelon state will again have a string tension component $KL$ but now the string fluctuation term which behaves as $\pi/(3L)$ for a bosonic string is unaffected by any Coulomb component. To extract the string fluctuation component most precisely, one should consider long strings so that the string model is most likely to apply. Furthermore the string model assumes that there is no restriction on the transverse modes of the string. Thus in lattice terms we need an $L \times S^2 \times T$ lattice with both $S$ and $T$ big. This has been looked at previously [7] but attention was drawn to the need to control the transverse size $S$ carefully in such studies. Here we pursue this study using new more accurate determinations of the torelon energies.

In order to determine the energy of the ground state torelon most accurately, we need to use sophisticated methods. These include building efficient operators on the lattice to create the torelon states by using fuzzing or smearing of the gluonic paths used. Using several such fuzzing levels also allows a variational method to be brought to bear on the energy extraction. We also need to obtain large statistics by using fast programs on powerful computers. Even so, we feel that it is worthwhile to study this problem using pure gauge $SU(2)$ as a first step, since simulation is much faster for the $SU(2)$ gauge group. This will still enable us to calibrate a string model in a non-Abelian gauge theory.

## 2  Torelon Energies from the Lattice

We wish to be able to study strings of length much greater that typical hadronic scales. At the same time we need to be in the parameter region where lattice results give dimensionless ratios independent of lattice spacing $a$. Choosing $\beta = 2.4$ allows us to explore a wide range of physical lattice sizes. This $\beta$ value is comfortably above the restoration of rotational invariance at $\beta \approx 2.2$ and is in the scaling region. With the scale for SU(2) given by $\sqrt{K} = 0.26/a = 0.44$ GeV at $\beta = 2.4$, the lattice spacing is $a \approx 0.6 \text{GeV}^{-1}$ so that a lattice of size $L = 20$ has physical size 2.4 fm. This means that we shall be able to study long strings.

Details of our lattice simulations are given in Table 1. For pure-gauge SU(2) colour, after equilibration using heatbath updates, we use an update consisting of several over-relaxation sweeps plus one heatbath sweep. The smaller lattice sizes were simulated using the Meiko Computing Surface at Liverpool (these lattices are labelled as having 1 path in Table 1). For the larger lattice sizes, where the higher mass of the torelon makes the calculation more error-prone, a more sophisticated analysis (using 3 paths) was employed and the CRAY YMP at RAL was used. For $L = 12$ we used both methods to cross check the results.

To create a torelon we use a colour flux loop encircling the periodic boundary condition in a spatial direction. Instead of just using a product of colour links in a straight line, we find that using iteratively fuzzed links is more effective. This models the wandering of the colour flux as it loops around the spatial direction, and hence improves the torelon operator so that it has a relatively large overlap with the ground state torelon. Based on previous work with static potentials [1], we use the following link fuzzing algorithm.

$$U_{\text{new}} = P_{\text{SU}(2)}(cU_{\text{straight}} + \sum_{1}^{4} U_{\text{staples}})$$

in which $P_{\text{SU}(2)}$ projects the resulting matrix back to SU(2). We find that $c = 2.0$ is effective and we use 15 iterative fuzzing levels. For some of the lattice sizes, three fuzzing levels are used so providing a basis for a variational analysis of the optimum combination for the ground state. We measure the zero-momentum correlation between such torelon operators at time separation $t$ on the lattice configurations.

The correlation between torelon operators behaves as $e^{-E(L)t}$ for time separation $t$ where $E(L)$ is the ground state torelon mass. However there is also a contribution from a torelon 'round the back' of the form $e^{-E(L)(T-t)}$. This will only be negligible if $e^{-E(L)T} \ll 1$. This requirement is the motivation for the time extent $T$ chosen.

To extract the ground state torelon energy for a given value of $L$ we evaluate the effective mass from correlations $W$ at adjacent $t$ values:

$$m_{\text{eff}}(t) = -\log(W(t)/W(t-1)).$$

The limit at $t \to \infty$ is then the required energy value $E(L)$. Since the relative error on $W$ increases with $t$, it is necessary to extract estimates of this asymptotic mass value from modest $t$ values. This is a reliable procedure because the approach to the asymptotic value of $m_{\text{eff}}$ is controlled by the comparatively large value of the energy gap $\Delta E$ between the ground state and the first excited state. This energy gap can be estimated within the string model and is expected to behave as $8\pi/L$ at large $L$.

In the somewhat similar case of the extraction of the ground state of the static potential $V(R)$, the corresponding string model estimate of the energy gap is $2\pi/R$. This estimate is confirmed by lattice measurements and indeed a more refined estimate valid at lower $R$ values has been suggested [1], namely
$$V(R)^2 = K^2 R^2 + 2\pi N K$$
for excited states where $N = 2$ is the first excited state with the same symmetries as the ground state ($N = 0$). This suggests that we use for torelons the expression
$$E(L)^2 = K^2 L^2 + 8\pi N K,$$
again with $N = 2$ contributing to the first relevant excited state, which gives $\Delta E = 8\pi/L$ in the limit $L \to \infty$. Evaluating this for $L = 20$ yields $\Delta E = E(N = 2) - E(N = 0) = 0.94$, compared to the value of 1.26 obtained from the expression $8\pi/L$. Thus even in this worst case of $L = 20$, we expect a rapid approach with increasing $t$ of the effective mass to a plateau as $e^{-\Delta E(L)t}$. Furthermore, as we discus below, we can measure $\Delta E$ using the variational method and we find 1.03 for $L = 20$ which agrees well with the theoretical expectations above.

To quantify the approach to a plateau we measured $m_{\text{eff}}(t) - m_{\text{eff}}(t+1)$ with correlated error evaluated by a bootstrap analysis of the blocks of measurements. When this difference is consistent with zero within its error we may assume that a plateau has been reached and use $m_{\text{eff}}(t)$ as an estimate of the asymptotic value required (this value is shown in bold type in Table 2). Since $m_{\text{eff}}(t)$ is a monotonically decreasing function of $t$, this estimate is strictly an upper bound. For smaller $L$ values, the plateau is clearly reached very quickly in $t$ and we have confidence in using the values obtained this way.

At larger $L$, the energy $E(L)$ is larger, so that the signal decreases faster with increasing $t$. Thus there is a less well-established plateau and we use several methods to estimate the asymptotic value and its associated error. We measure correlations between two torelon operators, each of which can have three different levels of 'fatness'. This gives a $3 \times 3$ matrix of correlations at each $t$ value. This extra information enables ground state and excited states to be separated more effectively.

One well established method [8] is to use a variational analysis to find the optimum basis for the ground state. This amounts to finding the eigenvector $u^0$ associated with the largest eigenvalue $\lambda^0$ of
$$W_{ij}(t) u_j^a = \lambda^a W_{ij}(t-1) u_j^a.$$

To minimise the effect of statistical errors, we choose $t = 1$ and then explore the $t$ dependence of that eigenmode to larger $t$:
$$C(t) = u_i^0 W_{ij}(t) u_j^0.$$
Then we define effective masses using ratios of $C(t)$ instead of $W(t)$. These results are those shown in the Table 2 for the 3-path cases. The variational matrix analysis above with $t = 1$ also gives an estimate for $\Delta E$ from $-\log(\lambda^1/\lambda^0)$ and this agrees well with the string expectation as mentioned above.

In principle this variational approach could completely remove any excited state contributions from $C(t)$. In practice, a statistically significant decrease with increasing $t$ of the effective mass obtained from $C(t)$ is seen. This comes from some residual contamination of excited states in $C(t)$. A way to obtain an estimate of the largest possible such contamination is to assume that it is all due to the first excited state with energy gap $\Delta E$ with the value of $\Delta E$ taken from the variational analysis of $t = 0$ to $t = 1$. Then one can extrapolate from the $0 \leq t \leq 2$ correlations $C(t)$ using a two exponential form with this prescribed energy difference $\Delta E$. The result of this analysis is presented in Table 2 as $t(2, 1, 0)$ which includes a full bootstrap error estimate. Since this method assumes that all the $t$ dependence of $C(t)$ is due to the *first* excited state, then it will provide a lower bound on the asymptotic mass value. Combining with the upper bound obtained from the plateau method, this gives a realistic systematic error for the $t$ extrapolation.

A third method is to fit the data directly to a two exponential form for the generalised Wilson loops themselves:
$$W_{ij}(t) = e^{-E(L)t}(b_i^0 b_j^0 + b_i^1 b_j^1 e^{-\Delta E\, t}).$$
Let us illustrate this by considering the case of $L = 20$ since that is the case when the plateau is least clear. Fitting $1 \leq t \leq 7$, there are 42 data points. This is too many data points for a reliable correlated $\chi^2$ fit [9]. Using instead an uncorrelated fit, we get $E^0(20) = 1.344(58)$ with $\Delta E(20) = 0.59(21)$. The fit is equally good when $\Delta E(L)$ is fixed at the value 1.03 obtained from the 0/1 variational determination: this fit gives $E^0(20) = 1.388(48)$. In the last column of Table 2, we give the results from these fits to $1 \leq t \leq 7$ with fixed $\Delta E(L)$. Since these fit determinations of $E(L)$ agree reasonably with the upper and lower bound estimates discussed above, we use these values in subsequent fits. For the case of $L = 12$, we find consistency within statistical errors between the small-$L$ method and the variational method as shown in Table 2.

It is known [7] that $E(L)$ can depend on transverse size $S$. There is a big effect for $L = 6$ as can be seen by comparing $S = 6$ [7] where $E(6) = 0.281(3)$ [7] with $S = 16$ where we find $E(6) = 0.164(5)$. However, $L = 6$ is close to the finite temperature phase transition at $L \approx 5.4$ where $E = 0$ for $S \to \infty$ while $E \neq 0$ for $S = L$. Thus we expect a strong dependence on $S$. In contrast, we find no dependence of $E(L)$ on $S$, provided $S \geq L$, if $L \geq 8$. See Table 2.

Previous work [7] obtained values at $\beta = 2.4$ for the torelon energy when $L = S$. A different fuzzing prescription (factor-of-two or Teper fuzzing) was used and only heatbath updates were performed. For $L = 16, 12$ and 8, the results of $1.037(33)$, $0.741(15)$ and $0.395(8)$ respectively agree with those we present here. The ground state overlaps are slightly higher with the fuzzing prescription we employ here compared to that used in [7].

## 3 String Models

The transverse size of the hadronic flux tube is expected to be larger than 0.5 fm. This corresponds, with our scale set as $a \approx 0.6$ GeV$^{-1}$, to about 4 lattice spacings. Thus any effective string theory should only apply for lengths greater that this transverse size. Thus we should focus on our determinations of the torelon energy $E(L)$ at larger $L$ to investigate possible string self energy contributions. Nevertheless, previous investigation [1] of the string with fixed ends from the static potential $V(R)$ found that a particular treatment of the effective string theory gave an improved description at smaller $R$ values. We use the same treatment here, so parametrising

$$E(L)^2 = (KL)^2 - \frac{2fK\pi}{3}$$

where $f = 1$ for a bosonic string. For large $L$, this gives the usual formula $E(L) = KL - f\pi/(3L)$.

Fitting $L$ from 8 to 20 using the data from the last column of Table 2, we obtain an acceptable fit with

$$K = 0.0688(4) \qquad f = 1.00(3) \qquad \chi^2 = 5.2$$

This fit is illustrated in figure 1. Clearly fixing $f = 1$ gives the same fit. More surprisingly, including the $L = 6$ datum gives essentially the same fit parameters.

Now since our fit expression is equivalent at large $L$ to

$$E(L)/L = K - \frac{f\pi}{3L^2}$$

we show the comparison of the fits with the data in figure 2 as a plot of $E(L)/L$ versus $L^{-2}$ to illustrate the straight line behaviour expected. The slope on the figure then is directly proportional to $f$. The continuous line shows the fit.

Our results give strong evidence for a term in $E(L)$ with $L^{-1}$ behaviour at large $L$. The presence of such a term is expected from a string self-energy contribution and so we compare the coefficient with that of simple string models. The coefficient $f$ is found to be completely consistent with the bosonic string value of $f = 1$. This determination of $f = 1$ is statistically significant when a fit is made for $L \geq 8$, but the data for $L \geq 12$ are also clearly consistent with $f = 1$ as can be seen from figure 2. It is clear that our results are inconsistent with $f = 0.25$.

In the case of string model fits to the string with fixed ends, one has to take into account the self energy of the static ends. Thus

$$V(R) = V_0 + KR - \frac{f\pi}{12R}$$

would be used at large $R$. This shows that the constant $V_0$ tends to obscure the contribution $\pi/(12R)$ coming from the self energy of the quantum fluctuations of the string. In the fits discussed above, we have assumed that a torelon has no such 'end effects'. Indeed there are no ends to a closed loop and the usual string model approach does not include such an end effect. If one takes an empirical view, however, an excellent fit ($\chi^2 = 0.12$) to our data for $L \geq 8$ can be made with $E(L) = 0.082(L - 3.09)$. This fit is the dotted straight line in figure 1, so it does not have any curvature towards the $L = 6$ point. It is possible to rule out this rather ad hoc model by requiring consistency between the string tension determined from torelons and from static potentials.

The static quark potential $V(R)$ at $\beta = 2.4$ has been measured. The string tension can be obtained from the limit of the slope $dV(R)/dR$ at large $R$, usually by making a Coulomb plus linear fit to $V(R)$. From $16^4$ lattices, the result is $K = 0.0728(6)$ [1]. From $24^4$ lattices [7], a fit from $R = 2$ to 12 yields $K = 0.0708(8)$, while a fit from $R = 3$ to 12 yields $K = 0.0705(16)$. Since the $V(R)$ values fitted are more accurately determined at smaller $R$, it is difficult to estimate the systematic error in extracting the string tension. One general feature is that the systematic errors in determining $V(R)$ tend to reduce the values at larger $R$ relatively more which in turn would reduce the value for $K$. Thus the determinations of $K$ from $V(R)$ tend to be over-estimates and we expect $K < 0.071$. Thus there is no discrepancy with the value $K = 0.0688(4)$ obtained from the conventional string model analysis of torelon energies. The value $K = 0.082$ from the model for torelons with $f = 0$ but an ad hoc self energy is inconsistent.

Thus for the conventional string model, there is no discrepancy between the string tension extracted from potentials and from torelons.

## 4 Conclusions

Confinement is viewed as arising through a colour flux tube having energy which increases with its length. By studying colour flux tubes of length $L$ up to 2.4 fm, we are able to confirm that a constant confining force (the string tension $K$) is present in a non-abelian SU(2) colour gauge theory. Such a constant confining force can be modelled by an effective string theory and it is interesting to probe the nature of this string theory.

We find that a careful study of the energy $E(L)$ of torelons (flux loops winding around the spatial boundary) can expose the possible self-energy contributions in a hadronic string. This approach has the advantage over a study of the static quark potential that no confusion arises from Coulombic contributions. We see a clear signal for a $1/L$ contribution to $E(L)$ as would arise from such a self energy contribution. The coefficient is close to that expected for the self energy of a bosonic string but is inconsistent with a Dirac string.

Our results are for pure gauge SU(2) colour fields, but we expect similar conclusions to apply for SU(3) colour fields.

Table 1: Lattice Simulations used

| Size | Updates | Configs measured | Number of paths |
|---|---|---|---|
| 6 $16^2$ 64 | 80000 | 1600 | 1 |
| 8 $8^2$ 32 | 209000 | 4180 | 1 |
| 8 $12^2$ 32 | 200000 | 4000 | 1 |
| 10 $10^2$ 32 | 600000 | 12000 | 1 |
| 12 $12^2$ 32 | 490000 | 9800 | 1 |
| 12 $12^2$ 32 | 80000 | 20000 | 3 |
| 16 $16^2$ 16 | 80000 | 20000 | 3 |
| 20 $20^2$ 20 | 64000 | 16000 | 3 |

Table 2: Torelon Energies

| $L\ S^2\ T$ | $t = 1/0$ | $t = 2/1$ | $t = 3/2$ | $t = 4/3$ | t(2,1,0) | $E(L)$ |
|---|---|---|---|---|---|---|
| 6 $16^2$ 64 | 0.175(4) | **0.164(5)** | 0.166(6) | 0.168(10) | | 0.164(5) |
| 8 $8^2$ 32 | 0.430(2) | **0.403(5)** | 0.403(7) | 0.400(11) | | |
| 8 $12^2$ 32 | 0.427(3) | 0.409(4) | **0.403(6)** | 0.406(13) | | 0.403(6) |
| 10 $10^2$ 32 | 0.620(4) | 0.574(5) | **0.564(7)** | 0.557(9) | | 0.564(7) |
| 12 $12^2$ 32 | 0.806(4) | 0.730(8) | **0.716(11)** | 0.715(25) | | |
| 12 $12^2$ 32 | 0.801(4) | 0.744(6) | **0.731(12)** | 0.730(23) | 0.717(7) | 0.732(18) |
| 16 $16^2$ 16 | 1.165(5) | 1.085(16) | **1.021(51)** | 1.045(129) | 1.042(23) | 1.060(29) |
| 20 $20^2$ 20 | 1.530(8) | **1.405(36)** | 1.512(152) | 0.825(477) | 1.336(55) | 1.388(48) |

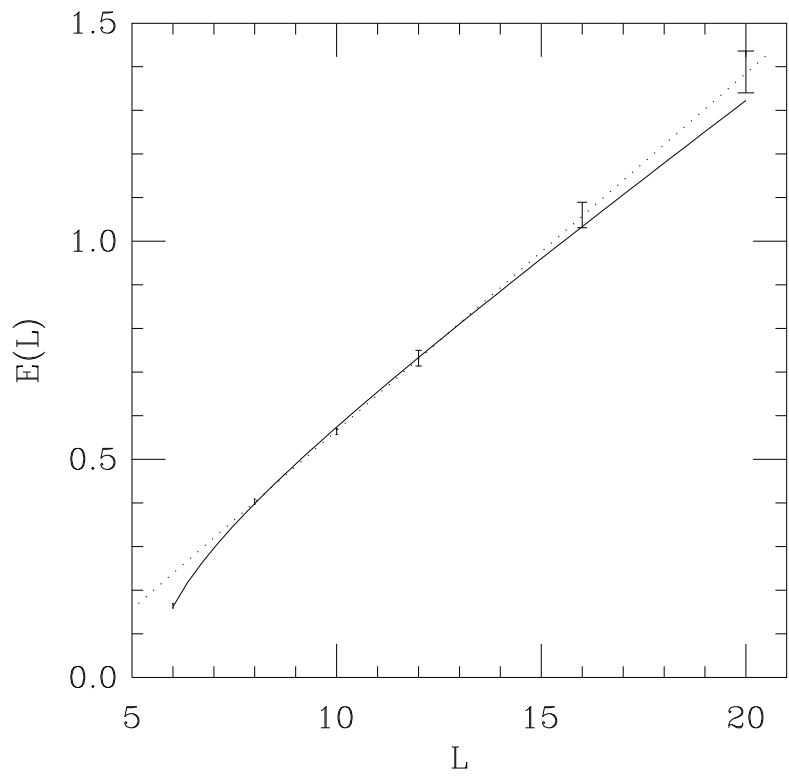

Figure 1: The torelon energy $E(L)$ versus $L$ for a periodic boundary of size $L$.

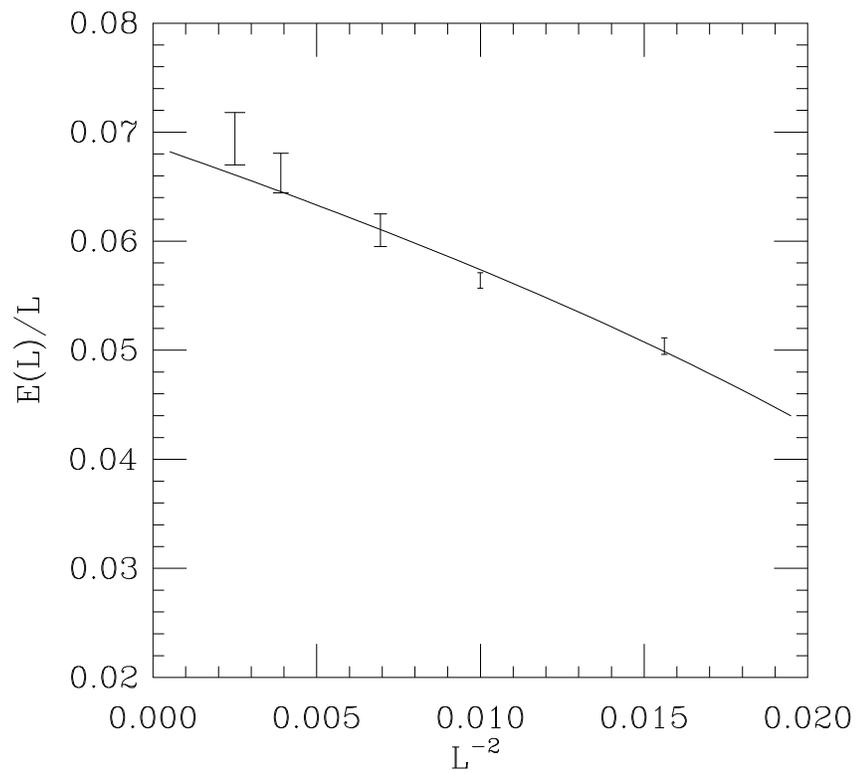

Figure 2: The torelon energy $E(L)/L$ versus $L^{-2}$ for a periodic boundary of size $L$.